\documentclass{npw}
\usepackage{graphicx}
\RequirePackage{xcolor}                  

\def\mbold#1{\mbox{\boldmath $#1$}}

\begin{document}

\title{ Nuclear tetrahedral states and high-spin states
        studied using quantum number projection method }
\author{
  S~Tagami, M~Shimada, Y~Fujioka, Y~R~Shimizu \\
  \it Department of Physics, Faculty of Sciences, \\
  \it Kyushu University Fukuoka 812-8581, Japan \\ and \\
      J~Dudek \\
  \it Institut Pluridisciplinaire Hubert Curien (IPHC), \\
  \it IN$_2$P$_3$-CNRS/Universit\'e de Strasbourg, F-67037 Strasbourg, France
}
\date{}
\maketitle

\begin{abstract}
We have recently developed an efficient method of performing the full
quantum number projection from the most general mean-field
(HFB type) wave functions including the angular momentum, parity as well as
the proton and neutron particle numbers.
With this method,
we have been investigating several nuclear structure mechanisms.
In this report, we discuss the obtained quantum rotational spectra
of the tetrahedral nuclear states formulating certain experimentally
verifiable criteria, of the high-spin states,
focussing on the wobbling- and chiral-bands,
and of the drip-line nuclei as illustrative examples.
\end{abstract}

\section{Introduction}

In the studies of the nuclear structure problems,
especially of the nuclear collective motion,
the underlying observables are the energy spectra
and the transition rates.
Although the mean-field methods, such as the Hartree-Fock (HF) or
the Hartree-Fock-Bogoliubov (HFB) method, or more recently,
the energy density functional (EDF) theories have been making
remarkable progress, nevertheless some further steps are necessary to obtain
the experiment-comparable information.
These involve in particular, the quantum number projection techniques,
which require even nowadays considerable numerical efforts.

We have developed recently an efficient method to perform
quantum number projections~\cite{TS12}.
In this contribution, we would like to show and discuss some results
of our recent studies which employ the projection method.
Ultimately, the framework of our approach is the standard one,
according to which the total wave function is described by
the linear combination of the projected states,
\begin{equation}
 |\Psi_{M;\alpha}^{INZ(\pm)}\rangle = \sum_{K,n} g_{Kn,\alpha}^{INZ(\pm)}\,
   \hat P_{MK}^I \hat P_{\pm}^{}\hat P^N \hat P^Z|\Phi_n^{}\rangle,
\label{eq:proj}
\end{equation}
where $|\Phi_n\rangle$ are the symmetry-broken HFB-type wave functions,
and $\hat P_{MK}^I$, $\hat P_{\pm}^{}$, and $\hat P^N \hat P^Z$
are the angular momentum, the parity,
and the neutron and proton number projectors, respectively.
The coefficients of the linear combination, $g_{Kn,\alpha}^{INZ(\pm)}$,
are obtained by solving the so-called Hill-Wheeler equation,
see e.g.~\cite{RStext} for the detailed formulation.
In the present work we only discuss the results with one HFB state,
i.e., without the configuration mixing of the generator coordinate method (GCM).

\section{Efficient method for projection}

The basic features of our method~\cite{TS12} for projection (and GCM)
are summarised as follows:
1) The most general symmetry-broken HFB-type
state can be treated including the breaking of time-reversal symmetry
(cranking), and the method can be applied not only to the even-even
but also to odd and odd-odd nuclei.
2) The harmonic oscillator basis is used at present,
but a new basis of the Gaussian expansion, which is more suitable for
the weakly bound system (halo/skin), is under development.
3) The truncation scheme based on the canonical basis is efficiently employed.
4) The Thouless amplitude $Z_{ll'}$ with respect to
the Slater determinant state is used,
\begin{equation}
 |\Phi\rangle=
 \exp{\Bigl(\sum_{ll'}Z_{ll'}^{}a^\dagger_l a^\dagger_{l'}\Bigr)}|\phi_0\rangle,
 \quad |\phi_0\rangle=\prod_{k=1}^N b^\dagger_k|0\rangle,
\label{eq:Thoul}
\end{equation}
where $N$ is the number of particles, and
$a^\dagger_k=b_k$ $(k \le N)$ or $a^\dagger_k=b^\dagger_k$ $(k > N)$
with $b^\dagger_k$ being the canonical basis creation operator
of HFB-type state $|\Phi\rangle$.
In this way, we can treat the situation of weak-pairing case
without any problem ($Z\rightarrow 0$), and can easily adopt
the Pfaffian formula~\cite{Rob} for calculating overlaps.

As for the Hamiltonian, we employ
the Woods-Saxon single-particle potential
and the schematic separable-type residual interactions.
For the particle-hole channel, we take the multipole interaction
($\lambda=2,3,4$),
whose form factor is the derivative of the Woods-Saxon potential,
with the selfconsistent value of the strength
as in \S6-5 of Ref.~\cite{BMtext} (no fitting at all).
For the pairing channel, we take the usual multipole-type
(the $r^\lambda$ form factor) with $\lambda=0$ and 2,
where the monopole (quadrupole) pairing strength
is fixed to reproduce the even-odd mass-differences (moment of inertia
of the ground state band), see Ref.\,\cite{TS12} for details.
Quite recently we have been able to perform the projection calculation 
making use of the more realistic Gogny force (the D1S parametrisation).
We will show a few preliminary results with it, while in most of
the calculations we will employ the Woods-Saxon potential with the schematic
multipole interaction schematised above (``WS+MI'').

Here we would like to emphasise the importance of the cranking term,
in our case, $-\omega_{\rm rot}J_x$, to generate the mean-field state
sufficiently rich in terms of various-symmetry states
as a `trial function to project from', especially for the ground state band.
In the axially symmetric ground state,
the mean-field wave function contains only $K=0$ component without cranking.
The cranking term induces the $K=\pm 1$ (time-odd) components,
which increase the moment of inertia considerably (by 30$-$40\%):
For this purpose, small cranking frequency is enough, e.g.,
$\hbar\omega_{\rm rot}$ of the order of a few 10's of keV,
and the results do not depend on a chosen value
of the frequency as long as it is sufficiently small;
see Fig.\,7 and~8 of Ref.\,\cite{TS12}.

\section{Tetrahedral nuclear states}

Our first application of the projection method will be to study
the tetrahedral-symmetry in certain nuclear states.
In the nuclear structure context the tetrahedral symmetry,
$T_d$, is sometimes referred to as {\em high-rank} point-group symmetry
to stress the existence of the 4-dimensional irreducible representation
of this group implying the four-fold degeneracy of certain nucleonic levels.
The latter mechanism leads to the extra nuclear stability
due to the specific shell effect (see below).

Tetrahedral shapes can be most easily described
as $\alpha_{32}$-deformation in the usual nuclear surface parameterisation
$R(\theta,\varphi)\propto
\bigl[1+\sum \alpha^*_{\lambda\mu}Y_{\lambda\mu}^{}(\theta,\varphi)\bigr]$.
When the single-particle orbitals, $\{\psi_n, e_n\}$,
are calculated in function of the tetrahedral deformation $\alpha_{32}$,
the existence of four-fold degenerate energy levels in addition to
the usual two-fold degenerate ones contributes to the appearance
of large shell gaps at certain nucleon numbers,
i.e., there exist ``tetrahedral closed shells'', see e.g.\,\cite{Dudek}.
The presence of the multipolarity $\lambda=3$ implies that
the parity is broken in the nuclear intrinsic frame of reference
and the presence of $Y_{32}$ in $R(\theta,\phi)$ above implies
that the axial-symmetry is broken.
Thus, with the cranking term, the single-particle Hamiltonian
has almost none of the otherwise often present mean-field symmetries,
and the general projection procedure of Eq.\,(\ref{eq:proj})
is indeed necessary.

The question arises: What are the specific properties of spectra generated
by tetrahedral-symmetric Hamiltonian?
We have recently performed the projection calculations
for the tetrahedral closed shell nuclei~\cite{TSD13}.
The tetrahedral shape is realised in molecular physics, e.g.,
in the methane (CH$_4$), the latter known as a tetrahedral rotor.
Paradoxically, it is a kind of ``spherical rotor'' in the sense that
all the three principal-axis moments of inertia are the same.
However, the group theory tells us that
only some specific spin-parity combinations are allowed in the quantum spectra
of the tetrahedral rotors according to irreducible representations
of the $T_d$ symmetry group. The ground state rotational sequence
of the tetrahedral double-closed shell nuclei belongs
to the so-called $A_1$ representation,
whose spin-parity combinations form the following sequence
\begin{equation}
  A_1:\begin{array}{l}
      0^+,\,3^-,\,4^+,\,6^+,\,6^-,\,7^-,\,8^+,\,9^+,\,9^-,\cr
      \,10^+,\,10^-,  \,11^-,\,2\times 12^+,\,12^-,\cdots.
    \end{array}
\label{eq:A1}
\end{equation}
Indeed we have shown that the results of the projections precisely
follow this property, see Fig.\,2 of~\cite{TSD13}.
Furthermore,  for small tetrahedral deformations the spectra
are approximately equidistant, resembling the multi-phonon structure
based on the $3^-$ phonon. This energy-spin dependence gradually
changes to the $E_I\sim I(I+1)$ type dependence,
when tetrahedral deformation increases, 
i.e., the transition from vibrational-like to rotational spectra
is clearly seen~\cite{TSD13}. 
\begin{figure}[t!]
\begin{center}
    \includegraphics[width=\columnwidth]{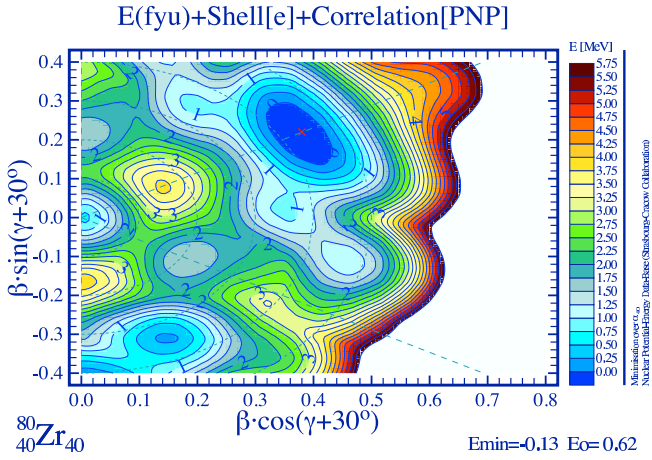}
    \includegraphics[width=\columnwidth]{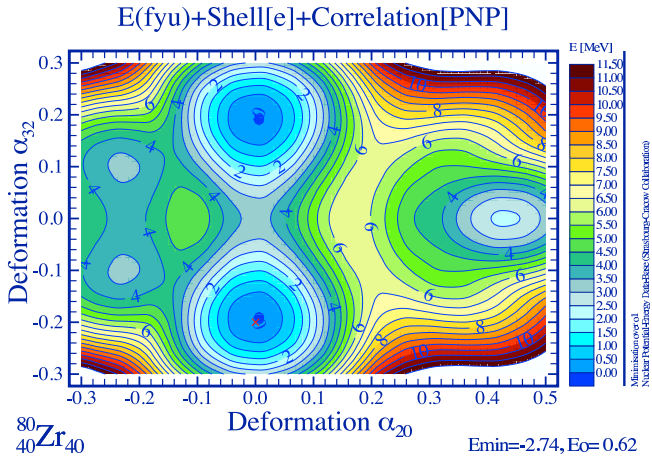}
\end{center}
\vspace*{-7mm}
\caption{\label{fig.01} 
According to the phenomenological mean-field calculations
with the Woods-Saxon Hamiltonian (universal parameterisation)
the `superdeformed' minimum, top, is the lowest in energy;
here at each $\{\beta,\gamma\}$ point the minimisation was performed
over $\alpha_{40}$. When the extra minimisation over
tetrahedral ($\alpha_{32}$) deformation is allowed, bottom,
the spherical minimum disappears, and the nucleus arrives
at the tetrahedral deformation minima with $\alpha_{32}\approx \pm 0.2$,
gaining 3.2 MeV, in an excellent agreement with our Gogny calculations.
[From \cite{JDM05}].}
\end{figure}

Among tetrahedral double-closed shell nuclei, $^{80}$Zr is
one of the most prominent.
Our recent Gogny D1S HFB calculations suggest
that the ground state has the tetrahedral deformation
$\alpha_{32}=0.11-0.12$, which has lower energy
than the prolate deformed minimum with $\beta\approx 0.45$
by more than 3 MeV.
Our phenomenological mean-field calculations with
the deformed Woods-Saxon potential, cf.~figure\,\ref{fig.01},
predict that within limited deformation space of
quadrupole and hexadecapole deformations,
the ground-state is expected to be very strongly deformed (`superdeformed')
with $\beta \approx 0.43$. However, when the minimisation
over the tetrahedral deformation is allowed,
the energy landscape changes dramatically,
the spherical minimum disappears and the tetrahedral-symmetry minima
appear for $\alpha_{32}\approx\pm 0.2$, about 3.2 MeV below. 

The doubly-magic tetrahedral nucleus $^{80}_{40}$Zr$^{}_{40}$ is
a very exotic one. Whereas experimentally only one rotational band is known,
with clearly identified $E2$-transitions, no experimental data exist
for the lighter isotopes. Given the fact
that {\em strictly tetrahedral-symmetric} configurations generate
neither quadrupole nor dipole moments,
the population and observation of tetrahedral rotational bands
belongs to the realm of `rare events', therefore most likely not yet seen.
For this reason the theory predictions from various independent calculations
and models indicating coherently that the ground-states in $^{80}$Zr
has tetrahedral symmetry deserves attention when planning new experiments.
As we believe, it is very likely that the observed `superdeformed' band
is an excited one, whereas the tetrahedral symmetric ground-state band
has never been seen. Since the tetrahedral configurations generate
in turn strong octupole moments, the future experiments should possibly
aim at this observable, if not in the very exotic $^{80}$Zr,
whose population may cause extra problems with the statistics,
then perhaps in some other isotopes in which similar properties are predicted.
\begin{figure}[ht]
\begin{center}
    \includegraphics[width=75mm]{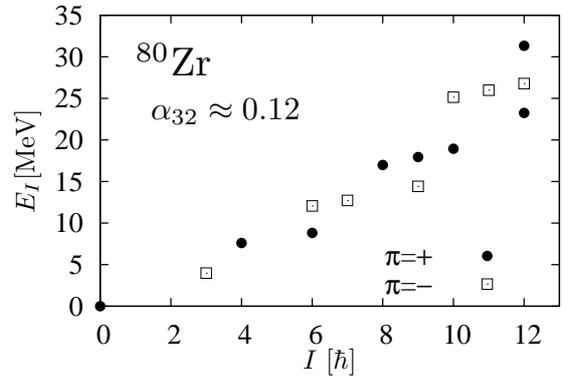}
\end{center}
\vspace*{-7mm}
\caption{\label{fig.02} 
Preliminary result of calculated excitation
spectrum for the tetrahedral ground state of $^{80}$Zr
calculated with the Gogny D1S interaction.}
\end{figure}

The pairing correlations are quenched for both
neutrons and protons because of the large shell gaps at $N=Z=40$.
In figure\,\ref{fig.02} we show our preliminary result of spectrum
with the Gogny D1S force as an example,
where the oscillator basis with $N_{\rm osc}^{\rm max}=12$ is used.
As it is clearly seen, the spectrum is `vibrational-like'
(approximately linear energy-vs.-spin dependence)
because of the rather small deformation
and follow the specific spin-parity combinations in Eq.\,(\ref{eq:A1}).
We have performed calculations with the WS+MI Hamiltonian
and obtained very similar results, what justifies the possible use of the
schematic interaction for describing the collective excitations.

The previous figure illustrates the results for a rather insignificant
deformation of the system of $\alpha_{32}\approx 0.12$.
To address the issue of the structure of the rotational bands
under the dominating presence of the tetrahedral symmetry,
let us consider first a relatively simple case: The nucleus $^{81}$Zr,
in which one neutron is added on top of the tetrahedral-closed shell
$N=40$ in $^{80}$Zr. To guarantee that tetrahedral symmetry clearly
dominates let us consider rather extreme deformation of $\alpha_{32}=0.4$.

\begin{figure}[ht]
\begin{center}
\includegraphics[width=75mm]{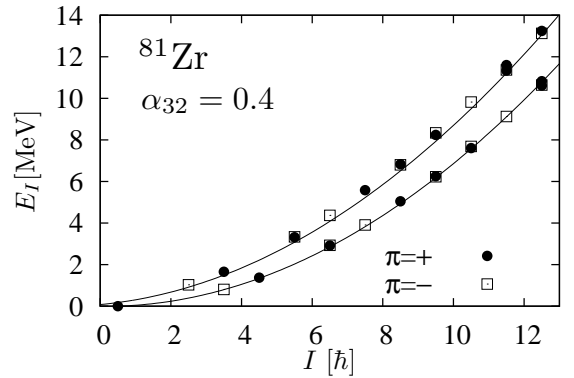}
\end{center}
\vspace*{-7mm}
\caption{\label{fig.03}
Calculated spectrum of $^{81}$Zr with the WS+MI Hamiltonian;
the odd neutron was placed at one of two-fold $E_{1/2}$ levels
near the Fermi level.
The solid lines connect the energy levels of the rotor including
the effect of the Coriolis coupling, cf.~eq.\,(\ref{eq:Hpr}),
and surrounding text.}
\end{figure}

For the half-integer spins the $T_d$ group has
three irreducible representations, $E_{1/2}$, $E_{5/2}$ and $G_{3/2}$,
where $G_{3/2}$ is four-dimensional, while the others are two-dimensional.
In figure\,\ref{fig.03}, we show an example with 
the valence neutron occupying one of the two-fold degenerate states
corresponding to the $E_{1/2}$ representation.
It can be seen from the figure that the calculated excitation energies
form {\em two sequences} following the rule $E_I\sim I(I+1)$, 
thus interpreted as rotational, however, at the same time
the tetrahedral symmetry imposes specific `extra rules'.
These rules are sufficiently characteristic to help defining
the strategy of the experimental search of the tetrahedral symmetry
in this nucleus and will be discussed in some detail. 

The projected positive-parity states belonging to the lower branch
in figure\,\ref{fig.03}, form a spin-parity sequence
beginning with $I^\pi={1}/{2}^+$, there is no state with $I^\pi={5}/{2}^+$,
the next available state having $I^\pi={9}/{2}^+$.
From now on, the $\Delta I=2$ sequence is present with the states at 
$I^\pi={13}/{2}^+, {17}/{2}^+$, etc. The negative parity sequence
begins at the state $I^\pi={5}/{2}^-$,
the `expected' rotational band member at ${9}/{2}^-$ is missing,
and instead there are two negative parity states
present at $I^\pi={11}/{2}^-$ and $13/2^-$.
The first of them cannot be a band member in the usual sense
since there is no $E2$-transition possible,
neither populating nor de-exciting this state within the sequence. 

We may conclude that the lower energy branch contains
the positive parity band beginning with the band-head at $I^\pi=9/2^+$
and the negative-parity band beginning with the band-head
at $I^\pi=15/2^-$ only. Such bands could be mistaken with
what in a `common language' are called $K=9/2$ and $K=15/2$ bands.
Observe the symmetry imposed rule of $\Delta I =3$ `spin-shift'
between the two band-heads.
Similarly, the upper branch in figure\,\ref{fig.03} contains
$\Delta I=2$ positive parity sequence beginning with $I^\pi=7/2^+$
but the negative-parity sequence beginning only at $I^\pi=13/2^-$,
with again $\Delta I=3$ `spin shift' between the two.

There are yet more characteristic features of the rotational bands
in question imposed by the symmetry: Beginning with a certain spin value
some members of the band have nearly degenerate opposite parity partners
whereas some others have none, within the spin range considered.

Of course the above considerations should be accompanied by the discussion
of the characteristic features of the electromagnetic transition probabilities
whose branching ratios form another valuable set of criteria
for identifying the discussed symmetries in nature,
in analogy to the techniques developed in molecular physics.
However this part of the discussion is left for a forthcoming publication.

In order to understand the origin of the splitting
between the two sequences visible in figure\,\ref{fig.03},
we have considered the conventional particle-rotor model,
\begin{equation}
  H_{\mbox{\scriptsize p-rot }}
   = E_0+\frac{I(I+1)}{2{\cal J}}
   - \frac{\mbold{I}\cdot\mbold{j}}{\cal J}.
\label{eq:Hpr}
\end{equation}
In contrast to the axially-symmetric case, the $K$ quantum number is
not the good quantum number in the tetrahedral rotor case,
so that it is not totally trivial to evaluate
the Coriolis coupling in Eq.~(\ref{eq:Hpr}).

We represent the resultant rotor spectra including the Coriolis coupling
by the solid lines in figure\,\ref{fig.03},
where ${\cal J}$ is the average moment of inertia parameter
of the two projected sequences and $E_0$ is adjusted in such a way
that it shifts the lowest-energy state at zero energy.
The matrix elements of $\mbold{j}$ are calculated
with the Woods-Saxon single-particle states.
As it is seen from the figure, the splitting between
the two parabolic-type sequences can be interpreted as the result of
the Coriolis coupling.  Note that while the splitting caused by
the first order Coriolis coupling is only effective for the $K=1/2$-band
in the case of the usual axially-symmetric rotor, here
the splitting always appears because the coupling parameter is non-zero
due to the characteristic $K$-mixing of the tetrahedral rotor.
In the case of the $G_{3/2}$ four-fold degenerate level,
the pattern of the coupling is more complicated and
the spectra split into more than two sequences; in this case
numerical diagonalisation is necessary.

We may conclude that the mean-field calculations with the symmetry projection
predict the characteristic splitting of the lowest-lying rotational states
into two sequences as the ones in the figure.
The presence of these two sequences can be interpreted
as the result of the Coriolis effect in odd-A nuclei,
when the extra nucleon is added on top of the core
with the wave function belonging to the $A_1$ representation;
the details will be reported elsewhere.

For integer-spins the $T_d$ group has
five irreducible representations, $A_1,A_2$, $E$, and $F_1,F_2$,
for each of which characteristic spin-parity combinations
like the ones in Eq.\,(\ref{eq:A1}) are assigned,
see Appendix of Ref.\,\cite{TSD13}.
The HFB-type states, in which degenerate single-nucleonic levels are occupied
with the same probability like in the case of double-closed configuration
or of the completely paired configuration in the BCS treatment,
belong to the $A_1$ representation.

In order to analyse, to an extent schematically, the spectra belonging to
other irreducible representations, we have performed the following
construction: Assuming quenched pairing correlations,
we put two neutrons in a four-fold degenerate level,
belonging to the $G_{3/2}$ representation,
on top of the closed tetrahedral-magic shell in $^{80}$Zr at $N=40$.
In order to make the modelling as simple as possible,
we use the WS+MI Hamiltonian and choose for this `academic' test as before a
rather unrealistically large deformation of $\alpha_{32}=0.4$,
at which the tetrahedral-symmetry effects of the quantum rotor dominate.
\begin{figure}[ht]
\begin{center}
\includegraphics[width=75mm]{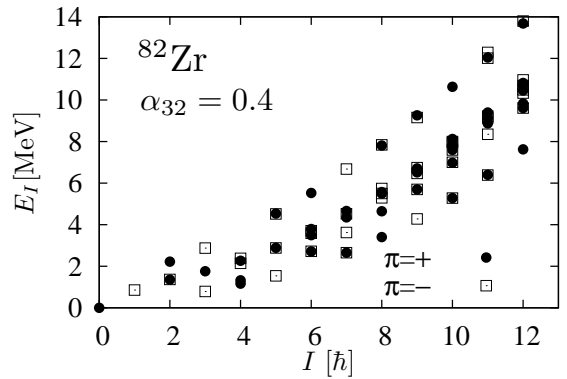}
\end{center}
\vspace*{-7mm}
\caption{\label{fig.04}
Low energy levels of $^{82}$Zr calculated with the WS+MI Hamiltonian,
where two neutrons occupy one of four-fold degenerate $G_{3/2}$ level
near the Fermi surface. }
\end{figure}

We show the resulting spectrum of $^{82}$Zr in figure\,\ref{fig.04}.
As can be seen, the energy sequences obtained follow
clearly the $E_I\sim I(I+1)$ proportionality.
The group theory tells us that the coupling of the anti-symmetric two-neutron
states of the $G_{3/2}$-symmetry can be decomposed according to the relation:
${\cal A}(G_{3/2}\times G_{3/2})=A_1+E+F_2$.
It then follows that the resulting energy levels, cf.~figure\,\ref{fig.04}
(details are omitted here and are left for the forthcoming publication)
belong to one of the three irreducible representations mentioned,
i.e., $A_1$, $E$, and $F_2$,
where the levels belonging to the three-dimensional $F_2$-representation
split further into three sequences.
The detailed analysis and interpretation of the results
in figure\,\ref{fig.04} can be constructed similarly
to the simpler case of $^{81}$Zr nucleus discussed above
and will not be presented here.

%------------------------------------------------
\section{High-spin states: Wobbling and Chiral rotations}

As the next example, we apply the projection method to the high-spin states
focussing on two symmetry-related types of rotational bands,
the wobbling- and the chiral-doublet bands.
The so-called wobbling bands arise within the quantum mechanical description
of the motion of the asymmetric top, see \S4-5 of Ref.~\cite{BMtext}.
In triaxially deformed nuclei the collective rotation about
all three principal axes are possible, what results
in the phonon-like multiple rotational bands built
on top of any single intrinsic configuration.
Such multiple band structures have been observed
in some of the so-called triaxial superdeformed (TSD) bands
in the Lu isotopes, see e.g.~Ref.\,\cite{Hag04}.

\begin{figure}[ht]
\begin{center}
\includegraphics[width=\columnwidth]{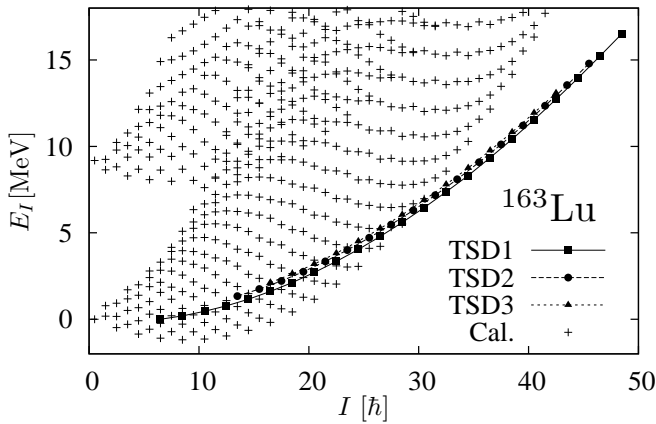}
\end{center}
\vspace*{-7mm}
\caption{\label{fig.05}
The results of our projection calculations with the WS+MI Hamiltonian, for the TSD bands
in $^{163}$Lu.
Calculated energy levels marked with the `$+$'-sign
are compared with the experimental results~\cite{Jen02} as indicated. }
\end{figure}

\begin{figure}[ht]
\begin{center}
\includegraphics[width=75mm]{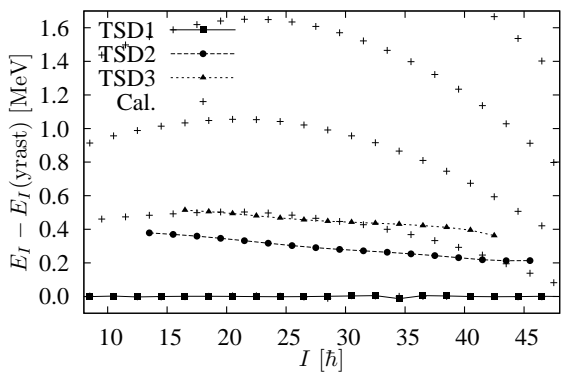}
\end{center}
\vspace*{-7mm}
\caption{\label{fig.06}
Relative spectra for the wobbling excitations in $^{163}$Lu. }
\end{figure}

\begin{figure}[ht]
\begin{center}
\includegraphics[width=75mm]{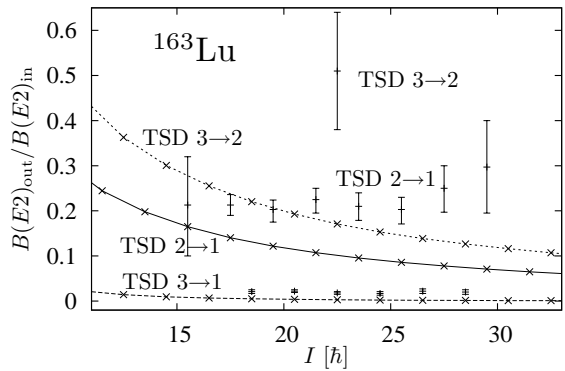}
\end{center}
\vspace*{-7mm}
\caption{\label{fig.07}
Calculated $B(E2)$ ratios
of the TSD bands in $^{163}$Lu are compared
with the experimental data~\cite{Jen02}
}
\end{figure}

We have performed the angular momentum projection calculations
for the TSD bands in $^{163}$Lu. In figure\,\ref{fig.05},
we show an example of the spectra calculated
with the WS+MI Hamiltonian, and in figure\,\ref{fig.06}
the corresponding relative excitation spectra.
In both figures, the experimentally observed
yrast, one-phonon, and two-phonon TSD bands
(TSD1, TSD2, and TSD3, respectively) are also included~\cite{Jen02}.
We also show the results of the out-of-band to in-band $B(E2)$-ratio
in figure\,\ref{fig.07}.

The mean-field deformation parameters used here are
$(\beta_2,\beta_4,\gamma)=(0.42,0.02,18^\circ)$.
They roughly correspond to those of the TSD minima
in the Woods-Saxon-Strutinsky calculation.
We have chosen 
the pairing gap parameters ${\mit\Delta}_n={\mit\Delta}_p=0.5$ MeV.
As for the cranking frequency to generate the HFB wave function,
we employed $\hbar\omega_{\rm rot}=0.2$ MeV,
associated with the $x$-axis cranking.
As it is seen from figure\,\ref{fig.05},
a regular multiple band structure appears,
however, the calculated moment of inertia of the yrast band is too small
compared with the observed one.  It should be emphasised that this kind
of multiple band structure never appears if the mean-field with too 
small a triaxiality.  

Concerning the results in figure\,\ref{fig.06},
let us note that the relative spectra are rather
sensitive to the mean-field parameters,
especially to the triaxiality $\gamma$
and the cranking parameter $\omega_{\rm rot}$.
The wobbling phonon excitation energy decreases when increasing $\gamma$,
while it increases when increasing $\omega_{\rm rot}$.
For example, we can reproduce approximately the one-phonon energies
either with $\gamma=30^\circ$ and $\hbar\omega_{\rm rot}=0.2$~MeV,
or with $\gamma=18^\circ$ and $\hbar\omega_{\rm rot}=0.1$~MeV,
although, in the latter case, the excitation energy decreases rapidly
and vanishes before spin 40$\hbar$.
On the other hand, the two-phonon excitation energies can be roughly
reproduced with $\gamma=35^\circ$ and $\hbar\omega_{\rm rot}=0.2$ MeV,
while we found it difficult to simultaneously reproduce both
the one-phonon and two-phonon spectra.

As for the $B(E2)$ ratios in figure\,\ref{fig.07},
the calculated values are smaller than the observed ones;
in order to reproduce the ratios larger $\gamma$ values are necessary.
The calculated $B(E2)$ ratio from the one-phonon to the yrast band
(TSD~$2\rightarrow 1$) decreases as a function of spin,
as long as the constant i.e.~spin-independent $\gamma$ is considered,
in contrast to the measured ratio that is almost constant or even increasing.
We would like to emphasise that the ultimate features
of the wobbling motion, obtained within the macroscopic rotor model,
are satisfactorily reproduced by the present fully-microscopic
projection calculation from a single intrinsic wave function,
although we are far from its satisfactory description at present. 

As another application investigated in this article let us present
the chiral rotation studied by employing again the projection method.
The chiral doublet bands are predicted for the triaxially deformed nuclei
with the proton and neutron angular momenta pointing to
two different directions~\cite{FM97}.
For example, in the triaxially deformed odd-odd nucleus with an odd proton
occupying the $\mbox{high-}j$ particle level and with an odd neutron occupying
the $\mbox{high-}j$ hole level, the three angular momentum vectors,
that of the proton, of the neutron, and of the collectively rotating nucleus,
tend to align with the largest, smallest, and middle principal axes,
respectively.

Under these circumstances, after Kelvin, the chirality mechanism arises
since the mirror image of an object (nuclear configurations
with the three vectors pointing to three directions in space)
cannot be superposed with the object itself through rotation.
Therefore to each left-handed combination of the three vectors
there should correspond a right-handed one with the result
that there should always exist two non-identical
although chiral-equivalent realisations of each discussed nuclear configuration
in a left-handed and right-handed version -- of the same energy.
However in the quantum systems one can imagine
a penetration of the potential barrier separating
the two associated potential energy minima
with the resulting energy-doublets split or degenerate,
depending on the properties of the potential separating the two minima.
This mechanism parallels the one  
of the parity doublet bands in pear-shape octupole-deformed nuclei.
\begin{figure}[ht]
\begin{center}
\includegraphics[width=\columnwidth]{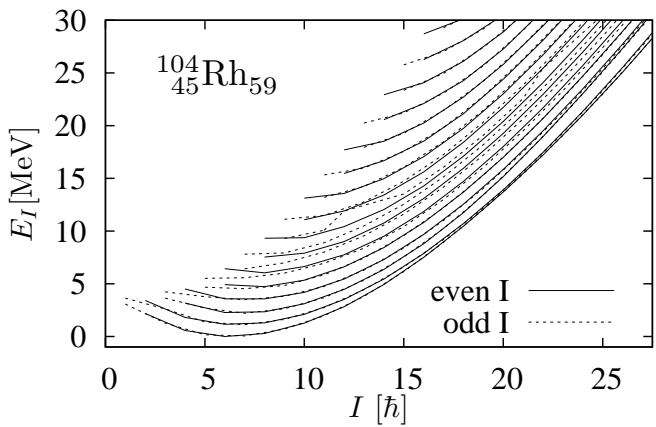}
\end{center}
\vspace*{-7mm}
\caption{\label{fig.08}
A result of projection calculation with the WS+MI Hamiltonian
for the chiral doublet band in $^{104}$Rh. }
\end{figure}

We wish to verify whether the chiral-doublet bands appear
in the calculations with the angular momentum projection
applied to the odd-odd nuclei.
An example of the results for the $^{104}$Rh nucleus
is shown in figure\,\ref{fig.08}.
In these calculations, we have fixed $\beta_2=0.3$ and $\omega_{\rm rot}=0$
together with $\beta_4=0$ and $\gamma=-30^\circ$ for simplicity.
The pairing gaps were self-consistently
calculated with the monopole pairing strengths selected to reproduce
the even-odd mass differences for both the neutrons and the protons.
Next we have studied the appearance/disappearance of the doublets of bands
by changing $\beta_2$ deformation and
the cranking parameter ($\omega_{\rm rot}$ about the $x$-axis).
We found out that the doublet-bands appear
within a rather restricted range of $\beta_2$,
whereas it seems to be more difficult
to obtain degenerate bands when increasing the cranking parameter.
As it can be seen from the figure, the odd- and even-spin members are
degenerate forming $|{\mit\Delta}I|=1$ bands
in the lower excitation energy region,
and when approaching at high-spins, $I\approx 15 - 30~\hbar$,
the two lowest bands produce doublet (rather than fully degenerate) bands,
which is expected for the chiral-symmetry breaking.

%------------------------------------------------
\section{Drip-line nuclei}

Finally we have studied rotational properties of certain unstable
drip-line nuclei. Our question to examine is:
How is the collective rotation affected
by the specific features of weak nucleonic binding,
the latter manifested through the skin and/or halo mechanisms.
To examine this issue, we have chosen the nucleus $^{40}$Mg
expected to lie close to-, or at the neutron drip-line.
According to our Gogny HFB calculations this nucleus
has a rather large axial deformation of $\beta_2\approx 0.34$
in spite of the fact that its neutron number is magic-spherical ($N=28$).
The density distribution indicates that there is a considerable
presence of the neutron skin and that the root-mean-square radius,
$\sqrt{\langle r^2 \rangle}=3.63$ fm, i.e.~about 15\% larger
than the that of other stable nuclei in the considered mass range.

\begin{figure}[ht]
\begin{center}
\includegraphics[width=70mm]{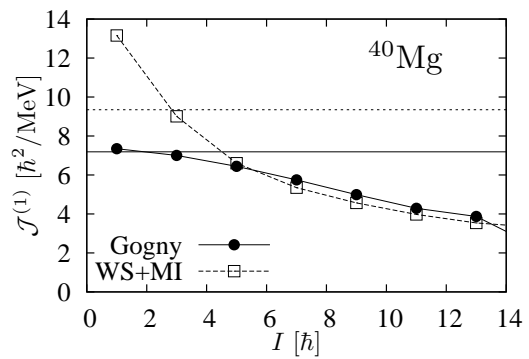}
\end{center}
\vspace*{-7mm}
\caption{\label{fig:Mg40moi}
Preliminary results for the moment of inertia defined as
${\cal J}^{(1)}\equiv (2I+1)/\bigl(E(I+1)-E(I-1)\bigr)$
of the ground state band in $^{40}$Mg.
The horizontal solid (dotted) lines denote
the rigid-body values with $r_0=1.2$ fm (with the calculated radius). }
\end{figure}

We have performed the angular momentum projection calculations
with both the WS+MI Hamiltonian and the Gogny D1S force.
For the WS+MI calculations, we use the deformation parameters
calculated by the HFB result with the Gogny D1S force.

Unfortunately, there are not enough experimental data to fix
the pairing force strengths in the WS+MI Hamiltonian for $^{40}$Mg,
so that we used in this paper the data from the stable isotope $^{24}$Mg
and made the extrapolation to fix the pairing strengths.
This procedure has ambiguities and the resulting spectra depend
on the specific choice of the stable isotope being used.
The energies of the first $2^+$ and $4^+$ states are
rather sensitive to this choice and may not be very reliable.
In figure\,\ref{fig:Mg40moi},
we show the moments of inertia obtained from the calculated
energy levels according to:
${\cal J}^{(1)}\equiv (2I+1)/\bigl(E(I+1)-E(I-1)\bigr)$.
We have used our two Hamiltonians, both of which give similar results
at increasing spins.
As it is shown in the figure, the moments of inertia
according to both calculations
decrease rather rapidly as spin increases.
This tendency is very rare for the inertia calculated
with the projection method, where the projection is performed from
the one HFB-type mean-field state.
We think that this feature of the decreasing moment of inertia
is specific for the weakly-bound nuclei.
Moreover, the value of inertia calculated with the Gogny D1S force
is considerably smaller than the rigid-body value even at the lowest spin.
This may indicate that the skin and/or halo-like neutron components,
whose density distributions tend to be spherical,
do not contribute to the moment of inertia of the collective rotation.

\section{Summary}

We have studied several subjects of contemporary interest in nuclear structure
by employing an efficient quantum number projection method,
which we have recently developed.
The importance of the cranking term is stressed to obtain
the correct magnitude of the moment of inertia
for the ground state rotational bands. 

After applying the projection onto the good angular-momentum
as well as parity, we have discussed the rotational band-structures
in the tetrahedral-symmetry states of $^{80,81,82}$Zr.
We have formulated some spectroscopic criteria which may be useful 
when proposing the experiments to test tetrahedral symmetry.
Although we have applied the group theory considerations to help understanding
the general spectroscopic features, the projection calculations are necessary
to obtain the numerical predictions.

Next, the results of the angular momentum projection
applied to the high-spin states have been presented.
Two kinds of symmetry-related rotational mechanisms,
the wobbling bands and the chiral-doublet bands were studied using
the fully-microscopic projection calculations.

Finally we have investigated the rotational spectra
of the weakly-bound drip-line nuclei,
where the effects of the halo and/or skin are expected to be important.
The results of our calculation suggest that
the moment of inertia of the ground state band
in such a drip-line nucleus may decrease with increasing spin,
which is quite different from those in the stable nuclei.

\begin{ack}
This work has been partly supported by Grant-in-Aid for Scientific Research (C)
No.~22540285 from Japan Society for the Promotion of Science,
and by the Polish-French COPIN collaboration under project number 04-113.
\end{ack}


\begin{thebibliography}{99}
\bibitem{TS12} Tagami~S and Shimizu~Y~R,
 Prog.~Theor.~Phys.~{\bf 127}, 79 (2012).
\bibitem{RStext} Ring~P and Schuck~P,
 {\it The nuclear many-body problem} (Springer, New York, 1980).
\bibitem{Rob} Robledo~L~M, Phys.~Rev.~{\bf C\,79}, 201302(R) (2009).
\bibitem{BMtext} Bohr~A and Mottelson~B~R,
 {\it Nuclear structure}, Vol.\,II (W~A~Benjamin, Inc., 1975).
\bibitem{Dudek} J~Dudek at al.,
 Phys.~Rev.~Lett.~{\bf 88}, 252502 (2002);
 Phys.~Rev.~Lett.~{\bf 97}, 072501 (2006).
\bibitem{TSD13} Tagami~S, Shimizu~Y~R and Dudek~J,\\
 Phys.~Rev.~{\bf C\,87}, 054306 (2013).
\bibitem{JDM05} Dudek~J and Mazurek K, Nuclear Potential-Energy Data-Base
(Strasbourg Cracow Collaboration), 
{\small http://jacobi.ifj.edu.pl/$\sim$mazurek/static/index\_alm.php} 
\bibitem{Hag04} Hagemann~G,
 Eur.~Phys.~J.~{\bf A\,20}, 183 (2004).
\bibitem{Jen02} Jensen ~D~R,
 Phys.~Rev.~Lett.~{\bf 89}, 142503 (2002).
\bibitem{FM97} Frauendorf~S and Meng~J,
 Nucl.~Phys.~{\bf A\,617}, 131 (1997).
\end{thebibliography}
\end{document}